\documentclass{PoS}

\def \inte {$INTEGRAL$}
\def \xmm {$XMM$-$Newton$}

\def \src {IGR\,J16418--4532}

\def \hcm {\hbox {\ifmmode $ atom cm$^{-2}\else atom cm$^{-2}$\fi}}

\def \mnras {MNRAS}

\title{Supergiant Fast X--ray Transients as transient sources in High Mass X--ray Binaries}

\ShortTitle{SFXTs as transient sources in HMXBs}

\author{\speaker{L.\ Sidoli} \thanks{Invited talk}\\
        INAF, Istituto di Astrofisica Spaziale e Fisica Cosmica, \\
         Via E.\ Bassini 15,   I-20133 Milano,  Italy\\
        E-mail: \email{sidoli@iasf-milano.inaf.it}}

\abstract{Supergiant Fast X--ray Transients  (SFXTs)
represent the most extreme case of X--ray variability in 
High Mass X-ray Binaries hosting blue supergiant companions.
Mainly discovered thanks to the INTEGRAL monitoring of the Galactic plane, 
these hard X-ray transients display dynamic ranges which can span five orders of magnitude.
This intensity variability is associated with accretion onto Galactic compact objects
(mostly neutron stars) through a physical mechanism which is still poorly understood. 
A review of the current status of our understanding on these sources is presented and discussed.
Finally, I present recent \xmm\ results about the SFXT \src, proposing that its variability behavior
is  due to a transitional accretion regime, intermediate between pure wind accretion and Roche Lobe Overflow.
This same regime could explain the X--ray activity of other ``intermediate'' SFXTs with narrow orbits.
}

\FullConference{The Extreme and Variable High Energy Sky\\
                 September 19-23, 2011\\
                 Chia Laguna (Cagliari) , Italy}

\begin{document}

\section{Supergiant Fast X--ray Transients: properties and hypotheses.}

Supergiant Fast X--ray Transients (SFXTs), a new class of X--ray sources
recognized thanks to INTEGRAL observations (Sguera et al. 2005, 2006; Negueruela et al. 2006),
represent the most extreme case of X--ray variability in 
High Mass X--ray Binaries (HMXBs) hosting early-type supergiants 
(Masetti et al. 2006; Pellizza et al. 2006; 
Nespoli et al. 2008; Rahoui et al. 2008; Chaty et al. 2008).
They show
dynamic ranges (ratios between the luminosity in outburst and in quiescence)
which can span from two (the so-called ``intermediate SFXTs'') to five orders of magnitude 
(observed, for example,
in the prototypical SFXT IGR~J17544--2619; 
Sunyaev et al. 2003, in't Zand 2005, Rampy et al. 2009).
This variability can be compared with a dynamic range of about one order of magnitude exhibited 
by supergiant HMXBs (SGXBs) with persistent X--ray emission, steadily accreting from
their massive donors and showing an accretion luminosity around $10^{36}$~erg~s$^{-1}$.
The SFXTs X--ray activity consists of outbursts lasting a few days (Romano et al. 2007, 
Sidoli et al. 2009a, Rampy et al. 2009), 
characterized by several short and bright flares, each with a duration between 
$\sim$10$^{2}$~s and $\sim$10$^{4}$~s,
reaching 10$^{36}$--10$^{37}$ erg s$^{-1}$ at the peak. 
One of the most sensitive and long observation of a SFXT was
performed by $Suzaku$  and was targeted on IGR~J17544--2619 (Rampy et al. 2009), the first
SFXT discovered by $INTEGRAL$ (Sunyaev et al. 2003). 
The observation lasted about 3 days, covering more than a half of the 
orbit (P$_{\rm orb}$=4.9~day, Clark et al. 2009), 
and displayed an extremely large dynamic range, with 
two phases of bright X--ray emission, which included the times of the periastron passage.

In these transients, both the duration and shape of the bright flares are variable, 
not only from source to source, but also for a given SFXT (e.g. Sidoli 2011). 
The SFXT extreme variability is driven by the accretion of matter from the supersonic 
wind of the supergiant companion, 
although the mechanism producing the transient emission is still not clear (see below).

X--ray spectra (0.1--100 keV) during flares are  well described either by a flat power law 
below 10 keV (Walter et al. 2006, Sidoli et al. 2006), with a photon index, $\Gamma$$\sim$0--1, 
and a high energy cut-off around 10--30~keV,
or by a bremsstrahlung model with a temperature, kT, of $\sim$15--40~keV. 
A more physical model could be applied to the less absorbed (10$^{21}$~cm$^{-2}$) 
SFXT, IGR~J08408--4503, allowing  recognition of a double component continuum during the bright flares,
with  a black-body together with a Comptonizing hot plasma (Sidoli et al. 2009b).

A low intensity flaring is usually present in the out-of-outburst state,
with an X--ray luminosity of 10$^{33}$--10$^{34}$~erg~s$^{-1}$.
In this state the X--ray spectrum is softer, with an absorbed 
power law with $\Gamma$$\sim$1--2  (Sidoli et al. 2008).
In deep observations with \xmm\ or $Suzaku$, a double-component model
is needed, with a hard power law together with a 
soft emission well described by a hot thermal 
plasma model (e.g. {\sc mekal} model in {\sc xspec}), with a temperature
of 0.2--0.3 keV, likely produced by the supergiant wind 
(Sidoli et al. 2008, 2010; Bozzo et al. 2010).
The power law spectrum present in the intensity state below 10$^{34}$~erg~s$^{-1}$
(and sometimes the detection of X--ray pulsations, e.g. Giunta et al. 2009)
demonstrates that the neutron star is still accreting, even outside the bright flares.
The lowest luminosity state at 10$^{32}$~erg~s$^{-1}$ is characterized by 
a very soft spectrum ($\Gamma$$\sim$6) and has been rarely observed 
(IGR~J17544--2619; in't Zand et al. 2005).

About a half of the members of the class are  X--ray pulsars, and they 
display very different spin periods, from 4.7~s (AX~J1841.0--0536, Bamba et al. 2001, 
although this periodicity
has been recently questioned by Bozzo et al. 2011) to 1210~s 
(IGR~J16418--4532, Walter et al. 2006, Sidoli et al. 2011).
A spin-phase spectral variability was observed and studied in detail
only in the SFXT IGR~J11215--5952 (Sidoli et al. 2007), the first member 
of the class known to display periodic outbursts (Sidoli et al. 2006)
every $\sim$165 days (Sidoli et al. 2007), likely
triggered near the periastron passage (although other binary system geometries are
also possible, see Fig.~9 in Sidoli et al. 2007).
The hardness ratio is indeed modulated on 
the pulsar spin period (P=187~s). 
The pulse profiles are different in two source intensity states, 
with a double-peaked shape during the bright flare, and a broad single pulse profile during 
the fainter state.
A combination of a power-law plus a blackbody component (kT $\sim $ 1-2 keV)
is a good deconvolution of the IGR~J11215--5952 continuum observed by \xmm,
resulting in a blackbody radius of  a few hundred meters, 
consistent with emission in the accretion column (e.g Becker \& Wolff 2005),
similar to what is often found in the X--ray emission produced by accreting Be/XRBs pulsars 
(e.g. La Palombara \& Mereghetti  2006).

\begin{figure}[ht!]
\begin{center}
\vspace{-0.5truecm}
\includegraphics*[angle=-90,scale=0.6]{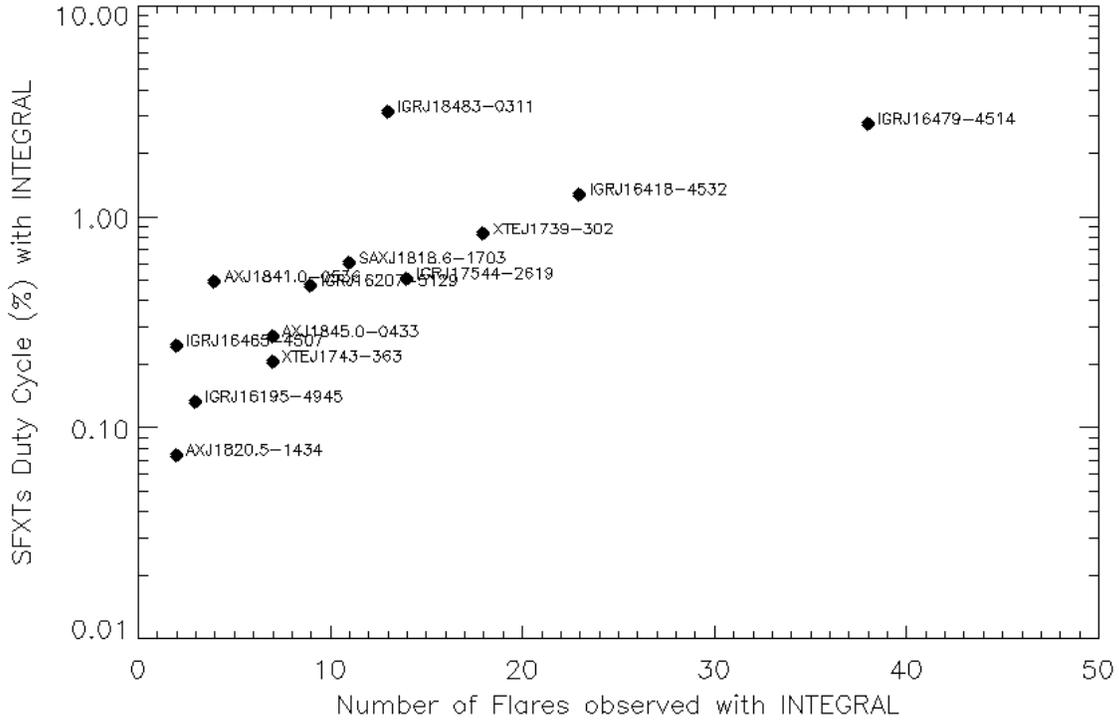}
\end{center}
\vspace{-0.75truecm}
\caption{\scriptsize Percentage of time spent in bright flares, as measured with \inte/IBIS, 
with respect to the total exposure
time of the source fields (data taken from Ducci, Sidoli, Paizis, 2010).
}
\label{lsfig:dc}
\end{figure}

The percentage of time spent by SFXTs in bright flaring activity (L$_{X}$$\sim$10$^{36}$~erg~s$^{-1}$) 
is lower than a few percent, 
and highly variable from source to source (see Fig.~\ref{lsfig:dc} for the duty
cycles observed by \inte/IBIS,
for SFXTs and candidate members located in the direction of the central regions of our Galaxy).
In the early times of these discoveries, SFXTs outbursts were 
apparently sporadic and unpredictable. 
On the other hand, a member of the class soon showed a periodicity
in the outburst recurrence (IGR~J11215--5952, Sidoli et al. 2006), as already mentioned above.
This SFXT seems to be a nice clock, since every time an X--ray satellite looks at it, at the predicted times
of its outbursts, it undergoes an outburst as expected. 
This 165~days periodicity, together with the shape of the X--ray light curve, is suggestive
of an orbital phase--locked steady structure, probably an equatorially enhanced wind component,
reminiscent of the Be--disks triggering the outbursts in 
transient Be/X--ray Binaries (Sidoli et al. 2007), although
we note that the structure of this supposed supergiant equatorial 
wind component could be very different from Be-disks, although it is 
able to trigger a source outburst 
every time the neutron star crosses it along its orbit.
A similar  wind component, inclined with respect to the orbital plane, 
is also very likely present in one of the better known SFXTs: XTE~J1739--302. 
It could indeed explain the presence of three peaks in the X--ray orbital light curve
(Drave et al. 2010; see also Bird 2011, these proceedings).

\begin{figure}[ht!]
\begin{center}
\vspace{+0.5truecm}
\includegraphics*[angle=0,scale=0.75]{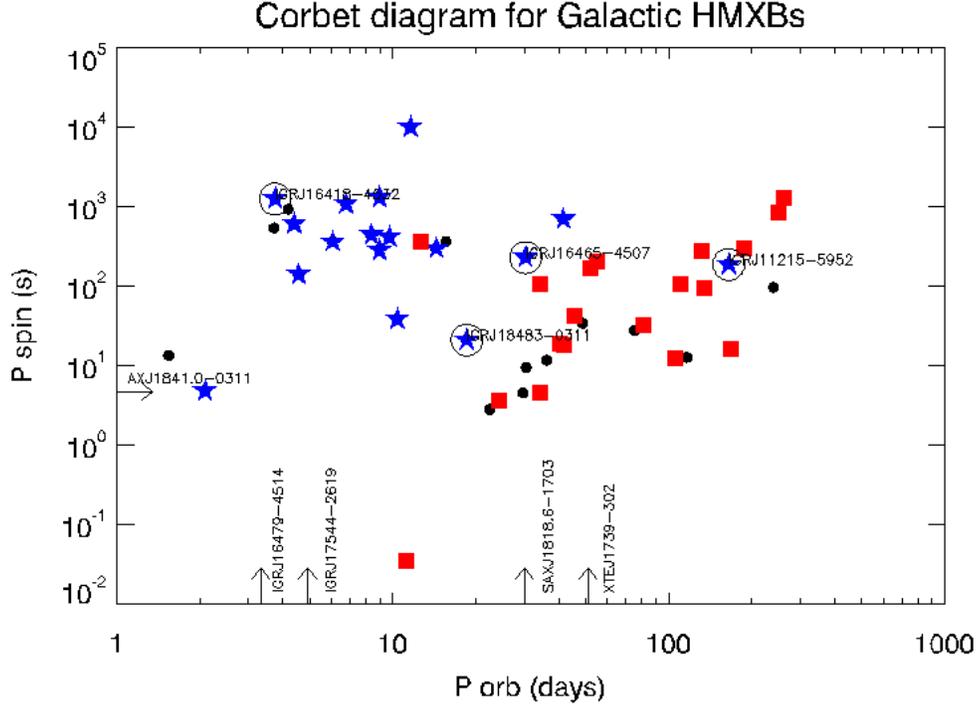}
\end{center}
\vspace{-0.75truecm}
\caption{\scriptsize Updated Corbet diagram of Galactic High Mass X--ray Binaries.
{\em  Blue stars} mark X--ray pulsars  with optically identified supergiant companions, while the
{\em  red squares} the sources
with optically identified Be donors (Liu et al., 2006). 
Supergiant Fast X--ray Transients are indicated by the {\em large circles} around blue stars.
Arrows mark SFXTs (or candidate SFXTs) where only one of the two periodicities are known.
}
\label{lsfig:corbet}
\end{figure}

The search for X--ray periodicities is a fundamental step towards the understanding of the
nature of these transient sources (Bird 2011, these proceedings). 
Long-term periodic X--ray modulations were later 
found in a few SFXTs, thanks to the timing analysis of large databases,
and interpreted as the orbital period of the binary systems.
Sometimes these orbital  modulations seem to be generated {\em mainly} or {\em exclusively}
by the outbursts emission (Sguera et al. 2007; Drave et al. 2010, 2011; Bird et al. 2009; Bozzo et al. 2009),
while in other cases (e.g. IGR~J17544--2619, Clark et al. 2009) the modulation is still present
even excluding the outbursts.
SFXTs orbital periods range from  3.3~days to 165~days and their positions in the Corbet diagram of 
spin period versus orbital period (Fig.~\ref{lsfig:corbet}) are strangely not confined 
only in the region typical of wind-fed accreting pulsars,
possibly suggestive either of an evolutionary link between SFXTs lying in the Be/XRBs region (or in the
bridge between wind-fed HMXBs and Be/XRB) and Be/XRBs (Liu et al. 2011), or of a similar
mechanism triggering the outbursts (as already discussed; Sidoli et al. 2007).

The SFXTs X--ray extreme variability is puzzling since these sources seem to be 
similar to SGXBs showing persistent X--ray emission: 
both are composed by compact objects (mostly neutron stars) accreting matter
from the  clumpy wind of the supergiant companion.
Thus, the main open issue deals with the difference between persistent SGXBs
and the SFXTs, 
where X--ray flares are usually attributed to the ingestion of single dense clumps
(in't Zand 2005, Walter et al. 2007, Negueruela et al. 2008, Ducci et al. 2009,
Rampy et al. 2009, Bozzo et al. 2011).
What drives the extreme transient behavior? If clumpy winds are ubiquitous in hot massive
stars (e.g. Sundqvist et al. 2010), why do SGXBs 
show steadily persistent X--ray emission 
while SFXTs display extreme X--ray transient activity?
Negueruela et al. (2008) suggested the important role of the orbital parameters (orbital separation
and/or eccentricity), but we now know that some SFXTs display even shorter orbital periods than persistent X--ray sources
(see, e.g., the Corbet diagram reported in Fig.\ref{lsfig:corbet}).
In the framework of clumpy winds, 
a  possibility could be that the wind porosity in SFXTs is intrinsically 
different from persistent systems, or that the clumping factor depends on the radial 
distance from the companion.

In close binaries, also the ionization effect by the X--ray source is thought to play a 
role (Ducci et al. 2010). 
In slowly rotating SFXT pulsars where quasi-spherical accretion takes place, 
the short X--ray flaring can  be produced by Rayleigh--Taylor instability 
(Postnov et al. 2011, these proceedings; Shakura et al. 2011).

A different structure (geometry) in the outflowing wind has been invoked by Sidoli et al. (2007), as discussed above,
suggesting a preferential plane for the outflow, inclined with respect to the orbital plane, thus triggering
the flaring activity only during the neutron star passage inside this wind component.

An alternative  explanation is that in SFXTs the accretion is inhibited for most of the
time by the presence of a centrifugal or a magnetic barrier produced by the neutron star
properties  (Bozzo et al. 2008), involving very high  magnetic fields (B$\sim$10$^{14}$--10$^{15}$~G) 
and slow spin periods  (P $\sim$1000~s).

\section{IGR~J16418--4532: a transitional accretion regime?}

\begin{figure*}
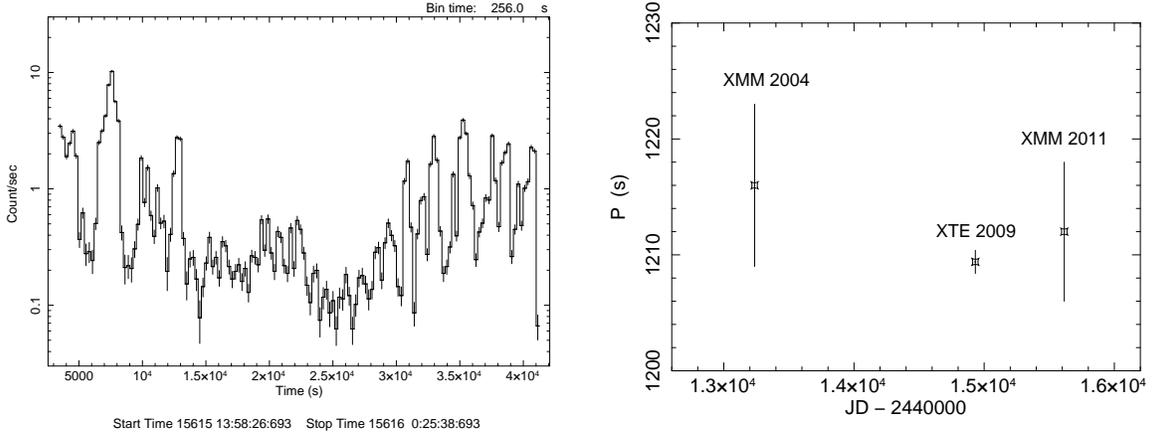

\centering
\begin{tabular}{cc}
\includegraphics[height=7.6cm, angle=-90]{ls_fig_3.ps} & 
\includegraphics[height=7.3cm, angle=-90]{ls_fig_4.ps} \\
\end{tabular}
\caption{\scriptsize  \emph{Left panel}: \xmm\ EPIC pn light curve (0.5--12 keV) of \src\ observed in February 
2011. 
\emph{Right panel}:  \src\ pulse period history. 
 }
\label{lsfig:igr}
\end{figure*}

\src\ is the SFXT with the narrowest orbit 
where both the orbital ($\sim$3.74~days;  Corbet et al. 2006, Levine et al. 2011) 
and the spin ($\sim$1200 s; Walter et al. 2006) periods are known (Fig.~\ref{lsfig:corbet}).
It belongs to the so-called ``intermediate'' SFXTs sub-class, where the observed dynamic range
is limited to two orders of magnitude.
We observed this transient in February 2011 with \xmm\ (Sidoli et al. 2011 for more details). 

In Fig.~\ref{lsfig:igr} (left panel) we show the source light curve observed with EPIC pn (0.5--12 keV),
while in the right panel the results of the timing analysis of our \xmm\ observation together
with two archival observations (\xmm\ in 2004 and $RXTE$ in 2009) is reported.

In the 2011 \xmm\ observation, the following interesting features are present: 
\begin{enumerate}

\item  a flaring activity with a dynamic range of two orders of magnitude, from
5$\times$$10^{34}$~erg~s$^{-1}$ to 5$\times$$10^{36}$~erg~s$^{-1}$ (assuming a distance of 13~kpc; 
Chaty et al. 2008; Rahoui et al. 2008);
\item  a hint of quasi-periodic flares (confined to the last part of the EPIC pn observation
reported in Fig.~\ref{lsfig:igr}), not related with the pulse period; 
\item  the pulsations, detected more clearly during 
the low luminosity state (unabsorbed flux level of 5$\times$$10^{-12}$~erg~s$^{-1}$), showed 
a double-peaked pulse profile, while a single peak was observed in 2004; 
\item  the EPIC spectrum did not show variability from flare to flare,
but displayed a lower absorbing column density in 2011 than in 
2004 (8.2($^{+0.8} _{-0.6}$)$\times$$10^{22}$~cm$^{-2}$ instead of 
18.6 ($^{+2.2} _{-2.3}$) $\times$$10^{22}$~cm$^{-2}$); 
the spectrum of the low intensity emission (central part of the
\xmm\ observation) is softer than during the flares;
\item  a soft excess is present in the cumulative flares spectrum, consistent with being produced by a photoionized wind
with a ionization parameter, $\xi$, of 125~$^{+60} _{-45}$~erg~cm~s$^{-1}$.

\end{enumerate}

Interestingly, the \src\ light curve observed in 2011 is very similar to
a simulated light curve performed by Blondin \& Owen (1997) of the accretion luminosity
in a HMXB undergoing a transitional accretion regime, intermediate between the 
full Roche Lobe overflow (RLO) and the pure wind accretion.  
According to these simulations, if the orbit is narrow (as in the case of \src) and the supergiant 
is close to filling its Roche Lobe (but it is not  undergoing RLO) 
the mass loss 
is  dominated by the strong outflowing
wind, but now with an additional contribution of a weak tidal gas
stream from the supergiant, focussed towards the neutron star. 
The dynamical interaction of the weak tidal
gas stream with the accretion bow shock around the compact object could produce 
the extreme variations in the accretion rate (Blondin, Stevens \& Kallman, 1991) 
shown in Fig. ~\ref{lsfig:igr} (left panel).

In this same scenario, the quasi-periodic flares
can be in principle produced by the strong asymmetry in the accreting wind, 
where temporary accretion disks of alternating directions could form (Blondin, Stevens \& Kallman, 1991).

In conclusion, we propose that the X--ray variability (its high dynamic range on the
observed time scale), 
the narrow orbit and the quasi-periodic flaring activity,
all suggest that the X-ray emission from \src\
is driven by a transitional accretion regime, intermediate between
``pure'' wind accretion and ``full'' RLO.

We suggest  that this accretion regime
could explain both \src\ X--ray properties
and possibly other ``intermediate'' SFXTs with similarly short orbital periods.

\acknowledgments
I would like to thank the organizers, and especially Angela Bazzano, for their kind invitation 
at the workshop ``The Extreme and Variable High Energy Sky'', 
held in a very nice place like Chia Laguna (Cagliari, Italy) on September 19-23, 2011. 
This gave me the possibility to have very interesting discussions on SFXTs properties with
Angela Bazzano, Tony Bird, Konstantin Postnov and Vito Sguera.
This work was supported in Italy by ASI-INAF contracts I/033/10/0 and I/009/10/0, and by 
the grant from PRIN-INAF 2009, ``The transient X--ray sky: 
new classes of X--ray binaries containing neutron stars''
(PI: L. Sidoli).


\begin{scriptsize}

\end{scriptsize}

\end{document}